\documentclass[10pt,twocolumn]{iopart}
\usepackage{graphicx}
\usepackage{hyperref} 
\usepackage{epstopdf}
\usepackage[normalem]{ulem}
\usepackage[utf8]{inputenc}
\usepackage{float}
\usepackage{caption}
\usepackage{bm}
\usepackage{comment}
\usepackage{amssymb}
\usepackage{bbold}
\usepackage{braket}

\newcommand{\down}{\downarrow}
\newcommand{\up}{\uparrow}

\begin{document}
\title{State transfer in an inhomogeneous spin chain}

\author{M Iversen$^{1}$, R E Barfknecht$^{1,2}$, A Foerster$^2$ and N T Zinner$^{1,3}$}

\address{$^1$ Department of Physics and Astronomy, Aarhus University, Ny Munkegade 120, Denmark}

\address{$^2$ Instituto de F{\'i}sica, Universidade Federal do Rio Grande do Sul, Av. Bento Gon{\c c}alves 9500, Porto Alegre, RS, Brazil}

\address{$^3$ Aarhus Institute of Advanced Studies, Aarhus University, DK-8000 Aarhus C, Denmark}

\eads{\mailto{rafael.barfknecht@ufrgs.br}}

\date{\today}

\begin{abstract}
\noindent We present an analytical formulation for the study of state transfer in a spin chain in the presence of an inhomogeneous set of exchange coefficients. We initially consider the homogeneous case and describe our method to obtain the energy spectrum of the system. Under certain conditions, the state transfer time can be predicted by taking into account the gap between the two lowest energy eigenvalues. We then generalize our approach to the inhomogeneous case and show that including a barrier between two spins - which effectively reduces the numerical value of the coupling between them - can have the counterintuitive effect of reducing the state transfer time. We additionally extend our analysis to the case of multiple barriers. Our results may contribute to the understanding of spin transfer dynamics in long chains where connections between neighbouring spins can be manipulated.
\end{abstract}

\pacs{67.57.Lm, 75.76.+j, 67.85.-d}

\submitto{\jpb}
\maketitle

\section*{Introduction}
A reliable implementation of quantum communication \cite{nielsen2000quantum,Bennett2000,bose_prl,bose} depends on the understanding of quantum state transfer along discrete qubit systems. A simple way to realize such systems is to consider one-dimensional spin chains \cite{bose_2003,bose_2007,allcock_2009,Xiang_2010}, where the strength of the interactions between neighbouring sites is given by exchange coefficients, which also determine the time scale of spin dynamics. A major goal in this context is the possibility of transferring a given state from one site to another - usually along the whole length of a one-dimensional system - in an efficient way.

Experimentally, such models can be studied with different setups, such as quantum wires \cite{burgarth,Quay2010}, superconducting circuits \cite{romito,You2011,Petersson2012,nakosai} and optical waveguides \cite{Nemec2012,perez_leija}. From a theoretical standpoint, quantum state transfer and transport dynamics have been studied both in the cases of static \cite{nikolopoulos_2004,plenio_2004,christandl,avellino_2006_PRA,gualdi_2008_PRA,nikolopoulos_2004b,
chiara,man_hong,bellec,artem2,paganelli,marchukov,palaiodimopoulos,Behzadi2018} and dynamical \cite{petrosyan,murphy,artem3} models. 

An interesting perspective posed by some of these studies is to consider cold atomic setups as quantum simulators \cite{bloch1,Gross995} to produce one-dimensional spin systems in optical traps \cite{fukuhara,Hart2015,Hilker484}. In these systems, the hyperfine states of the atomic elements such as $^{6}$Li (for fermions \cite{jochim1,Hilker484}) and $^{87}$Rb (for bosons \cite{widera_2008_prl}) can act as the spin degree of freedom, and interactions can be manipulated with precision to optimize certain dynamical effects. By having a single atom in a different internal state, the dynamics of spin transfer can be studied in a highly controllable environment. These techniques can also be applied to study the related issue known as the impurity problem, where a single distinguishable particle moves around in a background of identical atoms. Such system have been shown to exhibit interesting dynamical effects, from polaron physics \cite{amin_polaron,artem_polaron,Grusdt_2017,tylutki} to Bloch oscillations \cite{gangardt_bloch} and quantum flutter \cite{Mathy2012}.

\begin{figure}[ht]
\centering
\includegraphics[width = 0.5\linewidth]{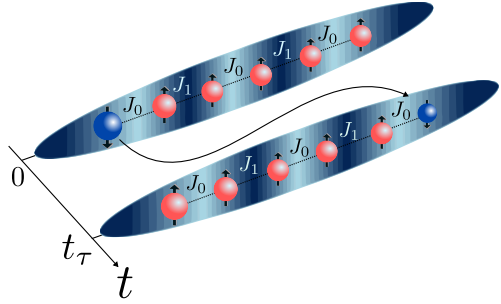}
\caption{We investigate the dynamics of a spin chain realized by a strongly interacting cold atomic system in an effectively one-dimensional geometry, where a single particle is considered to be in a different internal state (blue particle - flipped spin). We assume an inhomogeneous geometry where the presence of potential barriers can affect values of the exchange coefficients of the spin chain: for two neighbouring spins separated by a barrier (dark blue shaded region), the exchange coefficient is denoted by $J_1$ (and denoted by $J_0$ otherwise). The flipped spin is initialized at the left edge of the spin chain, and the total time required for the state transfer is denoted by $t_{\tau}$.}
\label{fig:1}
\end{figure}

In this work we present an analytical investigation of single-spin dynamics in a strongly interacting system, both in the cases of homogeneous and inhomogeneous trapping geometries (see Fig.~\ref{fig:1}). We interpret the inhomogeneous cases as spin chains that are split by potential barriers and show that, given certain conditions, the presence of such barriers can in fact enhance the transfer of a spin across the chain. Moreover, this effect does not require any dynamical control of the parameters. In the context of trapped cold atoms, it has been previously introduced in the strictly few-body regime, where double-well potential was taken into consideration \cite{barfk_josephson}. Here we present a more general approach for systems of arbitrary size and more than a single barrier. We additionally show that this problem can be explored analytically as well as with numerical simulations. We also take into account the possibility of having the initial state as a coherent superposition and the role of decoherence and potential imperfections on the dynamics. 

The geometry we consider is closely related to the Kronig-Penney model \cite{kronig_penney}, which has been shown to exhibit interesting static features, such as topological states and energy bands similar to the Hofstadter butterfly \cite{irina_kp}. Here, we focus on the regime of finite barriers and strong coupling away from the many-body limit. Quantum state transfer has also been studied in a configuration where the barriers are realized as local fields \cite{lorenzo}.  

Similar realizations may also be considered in condensed matter systems, for instance in a one-dimensional electronic system where the barriers would be realized by strongly localized heavy ions. A similar scenario has been considered for a bosonic gas in the presence of a barrier made by a single heavy ion \cite{gerritsma}, which could also be extended by taking into account a two-species system of atoms with highly imbalanced masses. 

\section{Analytics for an inhomogeneous chain}

\label{sec:DiagonalizingTheSpinHamiltonian}
We study the dynamics of a one-dimensional $XXZ$ chain of $N$ spins which is described by the following Hamiltonian:
\begin{eqnarray}\label{eq:SpinChainHamilton}
H_s = \mathbb{1} E_0 - A,
\end{eqnarray}
where $E_0$ is an energy constant and the operator $A$ is given by
\begin{eqnarray}
A = \sum_{i=1}^{N-1} J_i \left[ \frac{1}{2} (\mathbb{1}-\bm{\sigma}^i \cdot \bm{\sigma}^{i+1}) + \Delta(\mathbb{1}+\sigma_z^i \sigma_z^{i+1}) \right].
\label{eq:A}
\end{eqnarray}
In this expression $J_i$ is the position dependent nearest neighbour exchange coefficient and $\Delta$ is an anisotropy parameter that sets the strength imbalance for spins in the same state (in all our calculations, we assume $\hbar=1$). Let $\Omega_i$ and $\Lambda_i$ denote the eigenvalues of $H_s$ and $A$ respectively. To obtain the eigenvalues of $H_s$ it is sufficient to determine the eigenvalues of $A$ since they are related through 
	\begin{eqnarray}
	\Omega_i = E_0-\Lambda_i.
	\label{eq:EigenvalueRelation}
	\end{eqnarray} 
The problem of diagonalizing $H_s$ is therefore reduced to diagonalizing the operator $A$. We are now interested in obtaining an analytical expression for the matrix representation of $A-\mathbb{1} \Lambda$. To do so, we assume a basis of states where a single spin is flipped and write
\begin{eqnarray}
	\{ \ket{\down \up \ldots \up}, \ket{\up \down \ldots \up}, \ldots, \ket{\up \ldots \up \down} \}.
	\label{eq:Basis}
\end{eqnarray}
The form of the matrix representation depends now on the length of the spin chain, $N$, and the coupling constants, $J_i$. We will focus on the simple case of two distinct coupling constants across the chain. They are denoted by $J_0$ and $J_1$ with $J_0 \geq J_1$. Such systems could be realized by trapping ultracold atoms in an effective infinite well potential \cite{oelkers_2006} and separating the atoms by a number of finite barriers. 
If the atomic repulsion in the trapped system is strong enough, it has been established that the system can be described by a spin chain Hamiltonian \cite{artem,deuretz,xiaoling}. In this context, the exchange coefficients of the spin chain depend only on the geometry of the trapping potential (assuming a constant repulsive interaction), and $E_0$ is simply given by the total energy of a system of infinitely repulsive bosons (or spinless fermions). The coupling constant $J_1$ ($J_0$) would then correspond to the presence (absence) of a barrier between the atoms. Particles which are not separated by a barrier are assumed to be in the same ``well", and we will use $N_w$ to denote the number of wells in the system.

By assuming only a single flipped spin, the representation of $A- \mathbb{1} \Lambda$ is always given by a tridiagonal and symmetrical matrix. To simplify the notation, the following expressions are introduced:
\begin{eqnarray}\label{expressions}
&\beta = \frac{J_0}{J_1}-1, \nonumber \\
&\lambda = - \frac{\Lambda}{J_0} + 2 + 2\Delta \left(N - \frac{N_w}{1+1/\beta} -3 \right), \nonumber \\
&d_1 = 2 \Delta - 1 \nonumber \\
&d_2 = \frac{d_1}{1+1/\beta},
\end{eqnarray}
where $N$ is the length of the chain and $N_w$ is the number of wells. With these expressions $A-\mathbb{1} \Lambda$ is given by
\begin{eqnarray}\label{eq:A-1Lambda}
\fl
A-\mathbb{1}\Lambda = J_0\left( 
\begin{array}{cccccccccc}
\lambda + d_1 & -1 & & & & & & & \\
-1 & \lambda & -1 & & & & & \makebox(0,0){{\huge 0}}&\\
& -1 & \ddots & -1 & & & & & \\
& & -1 & \lambda & -1 & & & & \\
& & & -1 & \lambda + d_2 & -J_1/J_0 & & & \\
& & & & -J_1/J_0 & \lambda + d_2 & -1 & & \\
& & & & & -1 & \lambda & -1 & \\
& \makebox(0,0){\huge 0} & & & & & -1 & \ddots & -1 \\
& & & & & & & -1 & \lambda & -1 \\
& & & & & & & & -1&  \lambda + d_1
\end{array}
\right).\nonumber \\
\end{eqnarray}
The barrier size between two wells is now essentially determined by the parameter $\beta$: when $\beta = 0$ no barrier is present (we have a homogeneous system with single-valued exchange coefficients). When $\beta \gg 0$, we have a system which is split by an impenetrable barrier. In terms of the matrix structure, most elements in the diagonal, super- and subdiagonal are either $\lambda$ or $-1$. This is true except for the first row/column, last row/column and subspaces of the form
\begin{eqnarray}
\left(
\begin{array}{cc}
\lambda + d_2 & - J_1/J_0 \\
-J_1/J_0 & \lambda +d_2
\end{array}
\right),
\end{eqnarray}
where a barrier between the $i$'th and $(i+1)$'th sites of the chain is implied. This means that for each barrier in the system, there will be a corresponding subspace of this form in the matrix.

In order to find the eigenvalues belonging to $H_s$ we take the following approach: from equation~\ref{expressions} it is clear that $\Lambda$ is a function of $\lambda$. Therefore the values $\lambda_i$, which solve $\det(A-\Lambda \mathbb{1}) = 0$ are determined first. Next, these values are translated into the corresponding eigenvalues, $\Lambda_i$, belonging to $A$. Finally the eigenvalues, $\Lambda_i$, belonging to $A$ are used along with the relation $\Omega_i= E_0 - \Lambda_i$ to determine the eigenvalues, $\Omega_i$, belonging to $H_s$. For an arbitrary system, it is not expected that an analytical expression for the characteristic polynomial of the Hamiltonian exists. 
In this case, however, it is possible to obtain such an expression due to the particular structure of $A-\mathbb{1} \Lambda$. In \ref{sec:Appendix} we provide details on how to obtain this analytical expression. The resulting formula is
\begin{eqnarray}\label{eq:DetA}
\det(A- \mathbb{1} \Lambda) = (\lambda+d_1) \det(B_{N_w}) - \det(B_{N_w}'),
\end{eqnarray}
where the determinant of the matrices $B_{N_w}$ and $B_{N_w}'$ are given by the relation
\begin{eqnarray}\label{eq:DetBNw}
\left(
\begin{array}{c}
\det(B_{N_w}) \\ 
\det(B'_{N_w}) 
\end{array}
\right) = 
\left(
\begin{array}{c c} 
\gamma & \delta \\
\gamma'&  \delta'
\end{array}
\right)^{N_w-1}
\left(
\begin{array}{c} 
\det(B_1) \\ 
\det(B_1') 
\end{array}
\right).
\end{eqnarray}
In the expression above, $B_1$ and $B'_1$ are simple matrices while $\gamma$, $\gamma'$, $\delta$ and $\delta'$ are expressions which depend on $\lambda$. The eigenvalues of $A$ are found as the roots of equation \ref{eq:DetA}. The expression in equation \ref{eq:DetA} gets increasingly complicated for larger $N_w$. This can be seen from equation \ref{eq:DetBNw} where $(N_w-1)$ appears as a power of a matrix. This approach is therefore unfit to analyse systems consisting of many wells, $N_w \gg 1$. For increasing chain length, $N \gg 0$, on the other hand, the expression does not increase in complexity. This provides an interesting way to study state transfer also in long chains with a few inhomogeneities in the exchange coefficients.

\section{Applications to dynamics}
In this section we apply the method described above to find the eigenvalues for different spin chains and consequently analyse the dynamics of these systems. We start by comparing our approach to a two-level approximation, and then we study the dynamics in the presence of single and multiple barriers.

\subsection{Approximation as a two level system}\label{sec:ApproximationAsATwoLevelSystem} 
Consider a spin chain of length $N_\up + 1$, where we consider a single spin flipped. This particle represents an impurity in the system, while all other spins can be interpreted as a background of identical particles. Initially the homogenous system with no barriers is considered, which means all spin-spin interactions are the same and equal to $J_0$. We focus on comparing our results to a two-level approximation, where the two eigenstates with lowest energy are considered. The form of these eigenstates depends greatly on the strength of the anisotropy parameter $\Delta$. Here we take into account only the regime where $\Delta \geq 1$. For $\Delta = 1$, the background-background ($\uparrow\uparrow$) interactions and the background-impurity ($\uparrow\downarrow$) interactions are the same, which causes the impurity to be distributed evenly across the spin chain. In the opposite limit, $\Delta \gg 1$, a different behaviour is expected: since the background-background interactions are much weaker than the background-impurity interactions, the impurity will be located near the edges of the spin chain for the two eigenstates of lowest energy (these will have a similar structure but opposite parity). The background, on the other hand, will be located near the center of the spin chain. By initializing the impurity at one of the edges of the system we guarantee a large overlap of the initial wave function with the two states of lowest energy. Thus, we determine these two eigenstates, by direct diagonalization, to find 
\begin{eqnarray}
\ket{\psi_1} &= \frac{1}{\sqrt{2}} \big( \ket{\down \up \dots \up} + \ket{\up \dots \up \down} \big), \nonumber \\
\ket{\psi_2} &= \frac{1}{\sqrt{2}} \big( \ket{\down\up\dots\up} - \ket{\up \dots \up \down}  \big).
\label{eq:Eigenstates}
\end{eqnarray} 
As expected both eigenstates are linear combinations of basis elements where the impurity is at the edges of the spin chain. This behaviour can be qualitatively explained by considering the energy of the system. In the limit $\Delta \gg 1$ the background-impurity interactions are much stronger than the background-background interactions. This means that the system reaches the state of lowest energy if the impurity has the smallest possible interaction with the background, which is precisely the case of an impurity at the edges. If the system initially is a linear combination of $\ket{\down \up \dots \up}$ and $\ket{\up \dots \up \down}$ then it is described solely by the two lowest energy eigenstates, and the dynamics of such a system can then be approximated as that of a two level system. In this context, it is interesting to investigate the time it takes for the impurity to travel from one edge of the spin chain to the other, which we quantify as the transfer time $t_{\tau}$. When the system is initialized with the impurity at the left edge, $\ket{\down \up \dots \up}$, the state at $t=0$ is given by
\begin{eqnarray}
\ket{\Psi(t=0)} = \frac{1}{\sqrt{2}} \big( \ket{\psi_1} + \ket{\psi_2} \big).
\end{eqnarray}
The probability for the impurity of remaining at the left edge as the system evolves in time is calculated as the fidelity between $\ket{\Psi(t)}$ and $\ket{\down \up \dots \up}$. Likewise the fidelity between $\ket{\Psi(t)}$ and $\ket{\up \dots \up \down}$ is used as a measure of how well the impurity is transferred to the right edge. We denote these quantities by $F_l(t)$ and $F_r(t)$, respectively:
\begin{eqnarray}
F_l(t) &= |\braket{\Psi(t)|\down  \up  \dots  \up}|^2 = \frac{1}{2}\Big[ 1+\cos \Big(\frac{\Delta E}{\hbar} t \Big) \Big], \nonumber \\
F_r(t) &= |\braket{\Psi(t)|\up \dots \up \down}|^2 = \frac{1}{2}\Big[ 1-\cos \Big(\frac{\Delta E}{\hbar} t \Big) \Big],
\label{eq:Fidelity}
\end{eqnarray}
where $\Delta E = E_2 - E_1$ is the energy gap between the two eigenstates of lowest energy. The transfer time is obtained directly from equation \ref{eq:Fidelity} as half a period of $\cos(\Delta E t /\hbar)$
\begin{eqnarray}
t_{\tau} = \frac{\pi \hbar}{\Delta E}.
\label{eq:TransferTime}
\end{eqnarray}
From this equation it becomes clear that $t_{\tau}$ is reduced by increasing the energy gap between the two lowest energy eigenstates. It should be noted, however, that this analysis and equations \ref{eq:Fidelity} and \ref{eq:TransferTime} are only exact in the limit $\Delta \gg 1$. For smaller values of $\Delta$ the two lowest energy eigenstates will be linear combinations of all the basis elements. If $\Delta$ is sufficiently large the basis elements $\ket{\down \up \dots \up}$ and $\ket{\up \dots \up \down}$ will dominate and the two-level approximation will be more accurate, as we will show next.

\begin{figure}[ht]
\centering
\includegraphics[width = 0.5\linewidth]{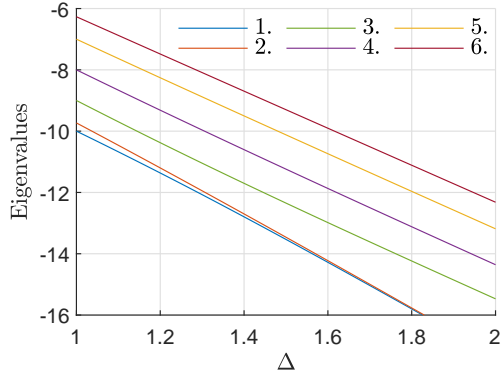}
\caption{The eigenvalues (in units of $J_0$) as a function of $\Delta$ for a spin chain of length $N=5+1$. Five sites are spin-up and one site is spin-down.}
\label{fig:2}
\end{figure}

We now consider the specific case of a small spin chain of length $N=5+1$. By analysing the energy spectrum, we can determine how accurately a two-level approximation describes the system dynamics. In this analysis only the energy difference between the eigenvalues are relevant, therefore the constant $E_0$ in equation \ref{eq:EigenvalueRelation} can safely be disregarded. The eigenvalues are determined for a series of different values of $\Delta$ and the result is illustrated in Fig.~\ref{fig:2}. Notice that, since we removed $E_0$, all energy values in this figure are negative.
	
For large values of $\Delta$, the energy gap between the first and second eigenstates is very small compared to all other cases. This behaviour confirms that the two eigenstates with lowest energy are completely isolated from the remaining eigenstates in the limit $\Delta \gg 1$. On the opposite end of the figure, where $\Delta = 1$, the background-background interactions and the background-impurity interactions are equal. The first and second eigenstates are therefore not expected to be isolated from the other eigenstates. In this regime, the energy gaps between all six eigenvalues are comparable. Fig.~\ref{fig:2} illustrates that the system is presumably well approximated as a two level system for large values of $\Delta$. 

\begin{figure}
\centering
\includegraphics[width = 0.5\linewidth]{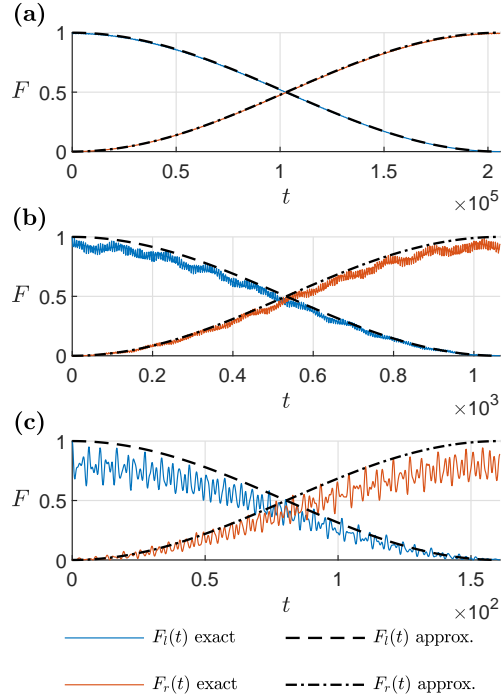}
\caption{Comparison between the exact fidelities (blue and red curves) to a two-level approximation (black dotted lines) for a spin chain with $N=5+1$. Initial state is given by $\ket{\down \up \dots \up}$. The anisotropy parameter is set as (a) $\Delta = 10$, (b) $\Delta = 3$ and (c) $\Delta = 2$. Time is given in units of $J_0^{-1}$.}
\label{fig:3}
\end{figure}

To determine how good this approximation actually is, the fidelities from equation \ref{eq:Fidelity} are compared to the exact fidelities. This comparison is shown in Fig.~\ref{fig:3}, where three different values of $\Delta$ are considered. The exact fidelities are determined by direct diagonalization of the Hamiltonian and using all eigenstates to calculate the time evolution of the system. For the largest value $\Delta$, the exact and approximated fidelities are almost identical, as expected from the previous discussion. In the other cases, $\Delta$ takes smaller values and the approximation is less accurate. By inspecting the figures it is clear that the local behaviour of the fidelity is less and less accurately captured by the approximation for decreasing $\Delta$. Despite this fact, the global behaviour is still well described by the approximation even for small values of $\Delta$. Therefore the two level approximation can be used to determine to what extend the impurity is located either at the left or right edge of the potential. 
The approximation can, however, not be used to determine how efficiently the impurity is transported to the other edge of the potential, since the approximation does not capture the local behaviour.

\subsection{Single barrier}
\label{sec:InsertingOneBarrier}
\begin{figure}[ht]
\centering
\includegraphics[width = 0.5\linewidth]{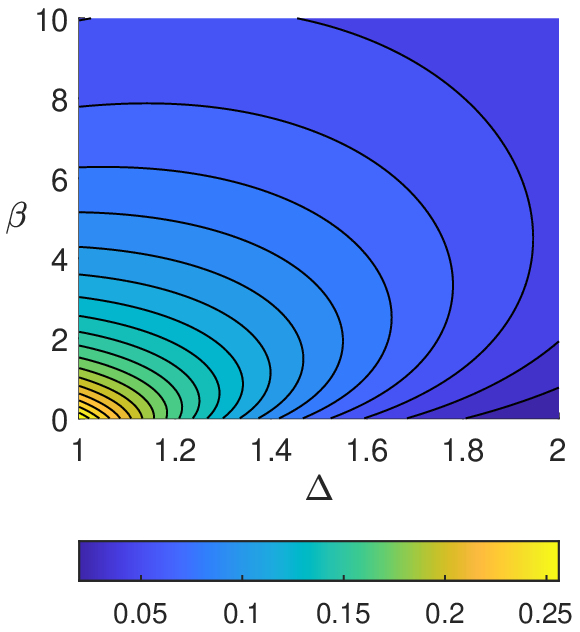}
\caption{Energy gap (in units of $J_0$) as a function of $\Delta$ and $\beta$ for a spin chain of length $N=5+1$ with a barrier between third and fourth particle. The figure shows that $\Delta E$ is monotonically decreasing as a function of $\Delta$ for a fixed $\beta$. For $\Delta > 1.2$ $\Delta E$ has a more complicated behaviour as a function of $\beta$. In this range of $\Delta$ there is a barrier size which maximizes the energy gap.}
\label{fig:4}
\end{figure}

We continue investigating the spin chain of length $N=5+1$ in the regime of $\Delta \geq 1$, only now a barrier is inserted between the third and fourth sites in the chain. As stated previously, the size of the barrier is controlled by the parameter $\beta = J_0/J_1-1$. Since the transfer time is inversely proportional to $\Delta E$, we aim to describe how the presence of this barrier affects the energy gap. Fig.~\ref{fig:4} shows the behaviour of $\Delta E$ as a function of $\Delta$ and $\beta$.

\begin{figure}[ht]
	\centering
	\includegraphics[width = 0.5\linewidth]{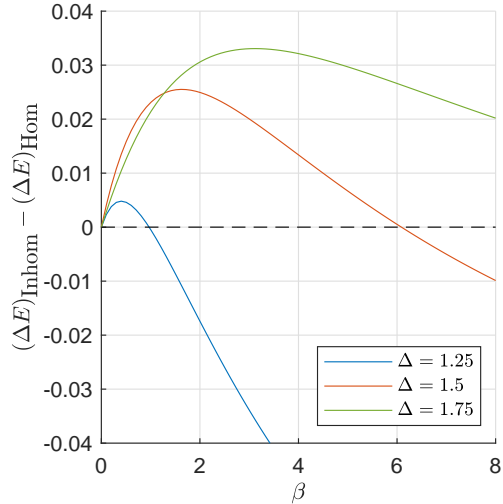}
	\caption{Difference in the energy gaps for the inhomogeneous 5+1 spin chain compared to the corresponding homogeneous system. $(\Delta E)_{\textnormal{\small Inhom}}-(\Delta E)_{\textnormal{\small Hom}}$ given in units of $J_0$ is shown as a function of the barrier size $\beta$. The figure illustrates that the energy gap can be increased by inserting a barrier in the center of the system.}
	\label{fig:5}
\end{figure}

We first investigate the behaviour of $\Delta E$ as a function of $\Delta$ for fixed $\beta$. From Fig.~\ref{fig:4}, it is seen that $\Delta E$ is monotonically decreasing as a function of $\Delta$ for all values of $\beta$. This behaviour can already be extracted from Fig.~\ref{fig:2}, where a small value of $\Delta$ resulted in a larger energy gap between the first and second eigenvalue while a larger value of $\Delta$ resulted in a smaller energy gap. 

Next, we investigate $\Delta E$ as a function of $\beta$ for a fixed $\Delta$. For small values of $\Delta$ (approx. $\Delta < 1.2$) the energy gap is monotonically decreasing as a function of $\beta$. This means that a smaller (larger) barrier results in a shorter (longer) transfer time. For larger values of $\Delta$ (approx. $\Delta > 1.2$), a different behaviour is observed. When a barrier is inserted, the energy gap increases, which corresponds to a decrease in transfer time. This means that the barrier enhances the motion of the impurity from one edge of the spin chain to the other.
The behaviour is surprising since the tunnelling of a single particle through a barrier is generally exponentially suppressed. To further investigate this effect, we focus of three cuts of Fig.~\ref{fig:4} corresponding to three different values of $\Delta$. In Fig.~\ref{fig:5} the energy gap is shown for the inhomogeneous system compared to the corresponding system where no barrier is present.

All three curves coincide at $\beta = 0$, since we have $J_1 = J_0$ and no barrier is present. Increasing the barrier size results in a larger energy gap until a maximum is reached. The behaviour of the energy gap can be explained in terms of the spin distribution in the chain with and without a barrier. As previously mentioned the background is primarily located at the center of the chain while the impurity is found at the edges. When a barrier is inserted the background is split, representing a smaller obstacle for the motion of the impurity. By comparing the three cases in Fig.~\ref{fig:5}, it is clear that the effect is greatest for large values of $\Delta$. Naturally, after a certain maximum value, increasing the barrier size will again suppress the motion of the impurity, even if we take into account the splitting of the background. 

\begin{figure}
	\centering
	\includegraphics[width = 0.5\linewidth]{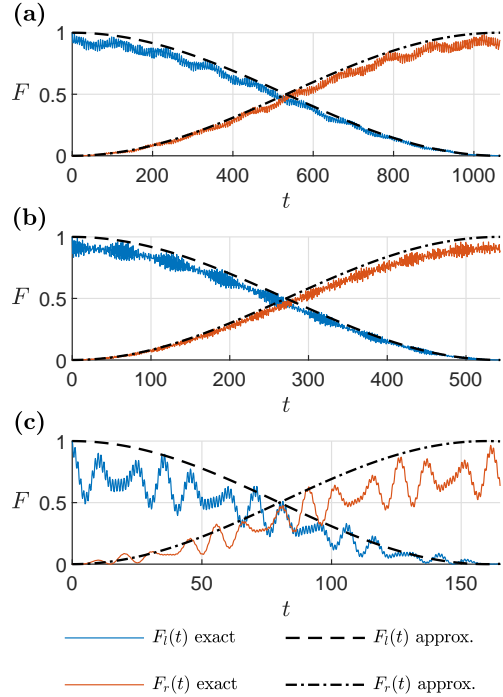}
	\caption{Comparison between the exact fidelities (blue and red curves) to a two-level approximation (black dotted lines) for a spin chain with $N=5+1$ and (a) $\beta = 0$, (b) $\beta = 1$, (c) $\beta = 9$. The anisotropy parameter is set as $\Delta = 3$. The presence of a barrier decreases the transfer time, which is given in units of $J_0^{-1}$.}
	\label{fig:6}
\end{figure}

The effect of inserting a barrier on the dynamics is more clearly seen in Fig.~\ref{fig:6}, where the fidelities $F_l$ and $F_r$ are shown as a function of time for different values of $\beta$ and $\Delta = 3$. In Fig.~\ref{fig:6} (a), where no barrier is present, the transfer time is $t_{\tau} \sim 1100 \, J_0^{-1}$, while in Fig.~\ref{fig:6}(c), where a large barrier has been inserted, the transfer time is reduced to $t_{\tau} \sim 160 \, J_0^{-1}$. This, however, results in a decrease in the absolute value of the transfer fidelity. Further studies may show how an optimal regime can be reached by demanding a minimum value for this quantity while also speeding up this process.

These results show that when a sufficiently small barrier is inserted into the system with large value of $\Delta$, the transfer time is reduced. An analogous effect for atoms in a harmonic double-well geometry has been obtained in \cite{barfk_josephson}. This indicates that the effect discussed here does not depend on the details of the trapping potential, given that a certain interaction regime is assumed. 

The effect described above is preserved when the initial state is a coherent superposition given by:
\begin{eqnarray}
\ket{\Psi(t=0)} = \frac{\ket \up + \ket \down}{\sqrt 2} \otimes \ket{\up \dots \up}.
\end{eqnarray}
The spin transfer is quantified by the fidelities $F_l(t) = |\braket{\Psi(t)|\down\up\dots \up}|^2$ and $F_r(t) = |\braket{\Psi(t)|\up \dots \up \down}|^2$. Considering $N=5+1$ particles in $N_w = 2$ wells and $\Delta = 3$ we observe dynamics very similar to the previous cases: inserting a barrier enhances the spin transfer across the chain. As expected, the maximum fidelity is limited to $F=0.5$ for the choice of initial state above. Thus we see that the conclusions hold even when the impurity is a coherent superposition.

\begin{figure}[ht]
	\centering
	\includegraphics[width = 0.5\linewidth]{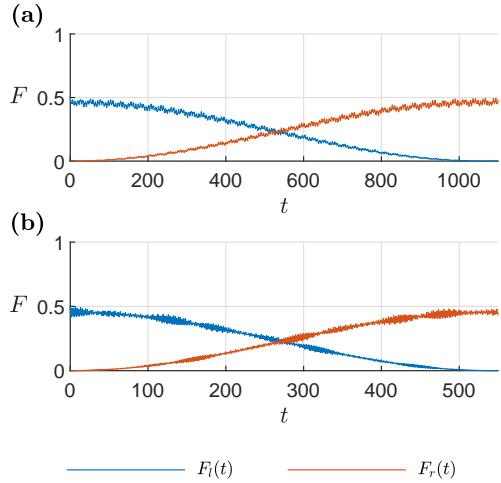}
	\caption{Impurity is a superposition of spin up and down, $(\ket \up + \ket \down)/\sqrt 2$.  The spin chain consists of $N=5+1$ particles distributed across $N_w=2$ wells with $\Delta =3$.  (a) When no barrier is present ($\beta = 0$) the impurity is transferred from the left to the right edge. (b) Inserting a barrier $\beta = 1$ enhances the spin transfer.}
	\label{fig:7}
\end{figure}

\subsection{Multiple barriers}
\begin{figure}[ht]
\centering
\includegraphics[width = 0.5\linewidth]{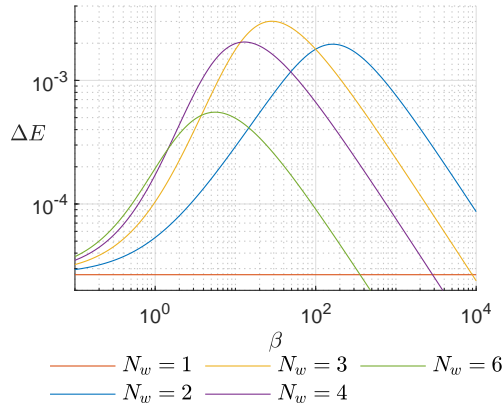}
\caption{A spin chain of length $N=11+1$ is split into $N_w = 1$, $2$, $3$, $4$ or $6$ wells. $\Delta E$ (in units of $J_0$) is illustrated as a function of $\beta$ for each value of $N_w$. In every case we assume $\Delta = 2$. In the figure all curves converge to $\Delta E = 0$ at $\beta \to \infty$.}
\label{fig:8}
\end{figure}

In the previous section we showed that inserting one barrier can reduce the transfer time for an impurity moving across the chain. Here we investigate the possibility of a similar effect arising in a scenario with multiple barriers. For this purpose, a longer spin chain consisting of $N=11+1$ sites is considered. This particular number of particles is suitable for the study of multiple barriers since it can be split into $N_w = 1$, $2$, $3$, $4$ or $6$ wells. Fig.~\ref{fig:8} shows $\Delta E$ as a function of $\beta$ for each of these number of wells, where $\Delta = 2$ is assumed. Notice that the red curve - the case where no barrier is present - is interpreted as the single-well homogeneous system.

For small values of $\beta$ (approx. $\beta < 1.4$), it is seen that the energy gap increases when the number of wells is increased. The transfer time is therefore reduced by inserting more barriers. This behaviour is explained by the same argument presented in section \ref{sec:InsertingOneBarrier}. The insertion of one or more barriers again results in a partial splitting of the background, which leads to a faster dynamics for the impurity. For a larger barrier size (approx. $\beta > 1.4 $) a more complicated behaviour is observed. In this region, more barriers do not necessarily lead to larger energy gaps. As an example, consider the curves with $N_w=4$ and $N_w=6$. These curves intersect each other at $\beta = 1.4$, which implies that separating the system into $N_w=4$ wells leads to a higher energy gap than $N_w=6$ for $\beta>1.4$. The same can be observed for other choices of $\beta$ and different $N_w$. Clearly, while splitting the background can lead to a smaller transfer time in some cases, for large barriers this effect quickly vanishes, which is expected from qualitatively arguments. In the regime of $\beta = 0$, all curves converge to the same value. This can be understood by the fact that there are no barriers present and all systems are effectively homogeneous. In the opposite end of the figure, $\beta \to \infty$, all curves go to $\Delta E=0$ except for $N_w = 1$. This behaviour is due to the fact that the barriers, in this regime, become impenetrable. The system is therefore degenerate and split into $N_w$ completely separated subsystems.

\section{The role of decoherence and imperfections}
Until this point we have assumed particles located in a spin chain isolated from the environment. In experimental settings, however, the exchange coefficients may not be perfectly fabricated and the system will have some uncontrolled interaction with its environment. In this section we investigate to what degree the results hold when the barriers are distributed unevenly and when the system experiences dissipation.

\subsection{Asymmetrical barriers}
We consider $N = 5+1$ particles distributed across $N_w = 3$ wells ($2$ barriers) with anisotropy parameter $\Delta = 3$ and barrier height $\beta=1$ . Ideally, the barriers are placed symmetrically, such that each well contains two particles. This is the setup already investigated, in which spin transfer is enhanced. If instead, the barriers are placed asymmetrically, we reach a different result. First, we place barriers between the third and fourth particles and fourth and fifth particles, such that there are two barriers in a row. In a second setup, we allow extra space between the barriers by placing them between the second and third particles and fifth and sixth particles. In both cases, we observe that the spin transfer is highly suppressed. This indicates that the type of dynamics described in the previous sections strongly depends on an even distribution of the barriers.
\begin{figure}[ht]
	\centering
	\includegraphics[width = 0.5\linewidth]{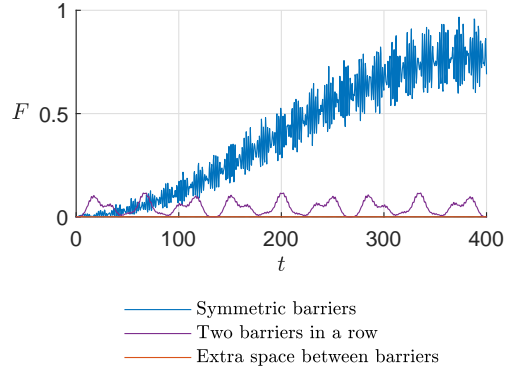}
	\caption{A spin chain of length $N=5+1$ with $\Delta = 3$ is distributed into $N_w=3$ wells by inserting $2$ barriers with $\beta=1$. Choosing a symmetrical barrier placement (blue) we observe spin transfer. Inserting two barriers in a row (purple) or with extra separation (red) destroys the spin transfer.}
	\label{fig:9}
\end{figure}
\subsection{Decoherence}
\begin{figure}[ht]
	\centering
	\includegraphics[width = 0.5\linewidth]{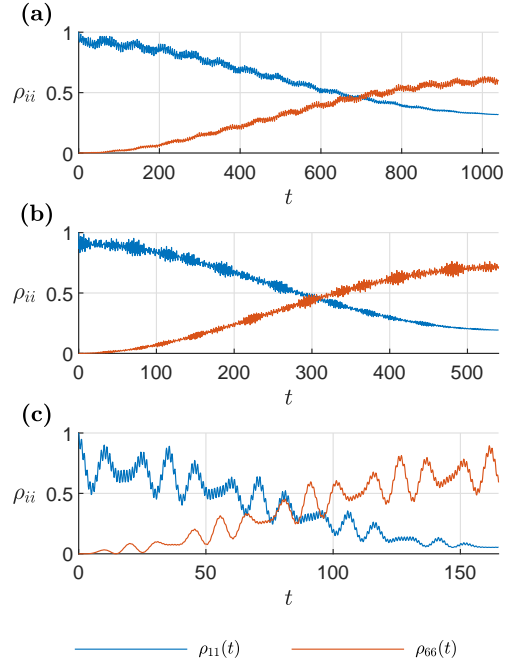}
	\caption{Spin chain of length $N=5+1$ separated into $N_w = 2$ wells with anisotropy parameter $\Delta=3$ and decay rate $\gamma=10^{-3} J_0$. Increasing the barrier size from (a) $\beta=0$ to (b) $\beta=1$ and (c) $\beta=9$ enhances the spin transfer from left to right edge of the spin chain.}
	\label{fig:10}
\end{figure}

Interactions between the system and its environment are treated by solving the Lindblad master equation \cite{Breuer}. For simplicity we assume the environment only causes transitions between the first and second system eigenstate. These transitions are modelled by a jump operator,
\begin{eqnarray}
L = \ket{\psi_1}\bra{\psi_2} + \ket{\psi_2} \bra{\psi_1},
\end{eqnarray}
with corresponding rate, $\gamma$. The impurity is initialized at the left edge, which corresponds to the density operator, $\rho(t=0) = \ket{\down \up \ldots \up} \bra{\down \up \ldots \up}$. Spin transfer to the right edge is described by the last diagonal element of the density operator, $\rho_{NN} = \braket{\up \ldots \up \down|\rho|\up \ldots \up \down}$. We consider $N=5+1$ particles in $N_w = 2$ wells with anisotropy parameter $\Delta = 3$ and decay rate $\gamma=10^{-3} J_0$. In figure \ref{fig:10}, we show that the dynamics is still enhanced in the presence of losses due to decoherence.

\section{Conclusions}
In this work we have studied the dynamics of a single flipped spin in a chain in the presence of inhomogeneities. This is done by including exchange coefficients with different numerical values, which can be interpreted as a homogeneous system split by potential barriers. We present an analytical procedure for finding the energy gaps in an arbitrarily large chain, which provides insight on the transfer dynamics of the impurity. We find that including a barrier in the center of the system can lead to an enhanced mobility of the flipped spin, an effect that can be generalized to cases with multiple barriers. We study different setups for varying numbers of particles and barriers. Additionally, we consider the role of initial states as coherent superpositions, decoherence and disorder in the barrier placement.
Our findings may contribute to the realization of optimized spin transfer in condensed matter and cold atomic systems. The latter may be an ideal candidate for this analysis due to the several possibilities regarding trap geometry manipulation and tuning of interactions. 

\ack
The authors thank Xiaoling Cui for interesting discussions on the model explored in this work. The following agencies - Danish Council for Independent Research DFF Natural Sciences, DFF Sapere Aude program, Carlsberg Foundation Distinguished Fellowship program, Conselho Nacional de Desenvolvimento Científico e Tecnológico (CNPq) and Coordenação de Aperfeiçoamento de Pessoal de Nível Superior (CAPES) - are gratefully acknowledged for financial support.

\appendix

\newpage
\section{Derivation of equation \ref{eq:DetA}}
\label{sec:Appendix}
In this Appendix we provide details on the analytical expression obtained for the inhomogeneous spin chain Hamiltonian. Consider $X$ to be a $n \times n$ matrix. We further define $Y$ as the following $(m+n) \times (m+n)$ matrix:
\begin{eqnarray}
Y = \left(
\begin{array}{ccccc}
\lambda  & -1 & 0 & \dots & 0 \\
-1 & \lambda & -1 & \dots & 0 \\
0 & -1 & \lambda & \dots & 0 \\
\vdots & \vdots & \vdots & \ddots &  \\
0 & 0 & 0 &  &  \scalebox{1.5}{$X$}
\end{array}
\right).
\label{eq:Y}
\end{eqnarray}
The entries in the first $m$ rows and $m$ columns consist solely of $\lambda$, $-1$ and $0$, while the entries in the last $n$ rows and $n$ columns are identical to $X$. 
The determinant of $Y$ is then given by
\begin{eqnarray}
\det(Y) = \frac{-\det(X')\sin(m\psi)+\det(X)\sin((m+1)\psi)}{\sin(\psi)},
\label{eq:DetY}
\end{eqnarray}
where $\psi$ is a complex number defined through the relation $\lambda = 2\cos(\psi)$. This result is useful since the matrix $A - \Lambda \mathbb{1}$ has rows of the form 
\begin{eqnarray}
	\Big(0,\dots,0,-1,\lambda,-1,0,\dots,0 \Big).
\end{eqnarray}
The goal now is to apply the result from equation \ref{eq:DetY} recursively to find $\det(A-\Lambda \mathbb{1})$. We start at the lower right corner of $A - \Lambda \mathbb{1}$. We further define $B_0$ and $B_0'$ as
\begin{eqnarray}
B_0 = 
\left(
\begin{array}{cc} 
\lambda & -1\\ 
-1 & \lambda+d_1 
\end{array}
\right),\,\,\,\,
B'_0 = \Big(\lambda+d_1 \Big),
\end{eqnarray}
where we point out that whenever a prime is added to a matrix it indicates that the first row and first column of that matrix has been removed. Analogously, we define
\begin{eqnarray}
B_1 = \left( \begin{array}{cccc}
\lambda & -1 & \cdots & 0 \\
-1 & \lambda & \cdots & 0 \\
\vdots & \vdots & \ddots & \\
0 & 0 & 0 & \scalebox{1.5}{$B_0$} \end{array}\right),
\end{eqnarray}
and let $B_1'$ be the matrix obtained by removing the first row and first column of $B_1$. Now we can replace $X$ by $B_0$ and $X'$ by $B_0'$ in equation \ref{eq:DetY} to determine $\det(B_1)$.
The next step is to define
\begin{eqnarray}
B_2 = \left( \begin{array}{ccccccc} 
\lambda & -1 & 0 & 0 & \cdots & &\cdots \\
-1 & \lambda & -1 & 0 & \cdots & & \cdots \\
0 & -1 & \ddots & & & & \\
0 & 0 & & \lambda + d_2 & -\beta & 0 & \cdots\\
\vdots & \vdots & & -\beta & \lambda +d_2 & -1 & \cdots\\
& & & 0 & -1 & & \\
\vdots & \vdots & & \vdots & \vdots & &\scalebox{1.5}{$B_1$}
\end{array}  \right),
\end{eqnarray}
and assume $B_2'$ is again the matrix obtained by removing the first row and column of $B_2$. This time the $X$ matrix in given by
\begin{eqnarray}
\left(
\begin{array}{cccc} 
\lambda +d_2 & - \beta & 0 & \cdots\\ 
-\beta  & \lambda + d_2 & -1 & \cdots \\ 
0 	& -1 &\\ 
\vdots & \vdots & & \scalebox{2}{$B_1$} 
\end{array}\right),
\label{eq:SecondStep}
\end{eqnarray}
and the matrix in equation \ref{eq:SecondStep} with first row and column removed is used as $X'$. Then equation \ref{eq:DetY} is again applied to find $\det(B_2)$. This step can be written in terms of the matrix multiplication
\begin{eqnarray}
\left(
\begin{array}{c} 
\det(B_2) \\ 
\det(B'_2) 
\end{array}
\right)
= 
\left(
\begin{array}{cc} 
\gamma & \delta \\ 
\gamma' & \delta' 
\end{array}
\right)
\left(
\begin{array}{c} 
\det(B_1) \\ 
\det(B_1') 
\end{array}
\right),
\end{eqnarray}
where the expressions $\gamma$, $\gamma'$, $\delta$ and $\delta'$ are given by
\begin{eqnarray}
\fl
\gamma = \frac{1}{\sin(\psi)}\Big\{\left[(\lambda+d_2)^2-\left(\frac{J_1}{J_0}\right)^2\right]\sin \left[\left(\frac{N}{N_w}-1\right)\psi \right]-\left(\lambda+d_2\right)\sin \left(\left(\frac{N}{N_w}-2\right)\psi\right)\Big\},\nonumber 
\\
\fl
\gamma' = \frac{1}{\sin(\psi)}\Big\{\left[\left(\lambda+d_2\right)^2-\left(\frac{J_1}{J_0}\right)^2\right]\sin \left[\left(\frac{N}{N_w}-2\right)\psi \right]-\left(\lambda+d_2\right)\sin \left[\left(\frac{N}{N_w}-3\right)\psi\right]\Big\},\nonumber
\\
\fl
\delta = \frac{1}{\sin(\psi)}\Big\{\sin\left[\left(\frac{N}{N_w}-2\right)\psi\right]-\left(\lambda+d_2\right)\sin\left[\left(\frac{N}{N_w}-1\right)\psi\right]\Big\},\nonumber 
\\
\fl
\delta' = \frac{1}{\sin(\psi)}\Big\{\sin\left[\left(\frac{N}{N_w}-3\right)\psi\right]-\left(\lambda+d_2\right)\sin\left[\left(\frac{N}{N_w}-2\right)\psi\right]\Big\}.
\end{eqnarray}
This step is then repeated $N_w-1$ times to obtain an expression for $\det(B_{N_w})$. This can again be written in terms of the matrix multiplication
\begin{eqnarray}
\left(
\begin{array}{c} 
\det(B_{N_w}) \\ 
\det(B'_{N_w}) 
\end{array}
\right)
= 
\left(
\begin{array}{cc} 
\gamma & \delta \\
\gamma' & \delta' 
\end{array}
\right)^{N_w-1}
\left(
\begin{array}{c} 
\det(B_1) \\ 
\det(B_1') 
\end{array}
\right).
\end{eqnarray}
Once $B_{N_w}$ and $B_{N_w}'$ are determined an expression for $\det(A - \Lambda I)$ can be obtained through the relation
\begin{eqnarray}
\det(A-\Lambda \mathbb{1} ) = (\lambda+d_1) \det(B_{N_w}) - \det(B_{N_w}').
\end{eqnarray}
These two equations yield an analytical expression for the characteristic polynomial of the operator $A-\Lambda \mathbb{1} $, which allows us to diagonalize the spin chain Hamiltonian. 

\section*{References}
\bibliography{bibliography}

\end{document}